\documentclass[conference,10pt]{IEEEtran}
\pdfoutput=1
\IEEEoverridecommandlockouts

\usepackage{cite}
\usepackage{amsmath,amssymb,amsfonts}
\usepackage{algorithmic}
\usepackage{multirow}
\usepackage{graphicx}
\usepackage{textcomp}
\usepackage{xcolor}
\def\BibTeX{{\rm B\kern-.05em{\sc i\kern-.025em b}\kern-.08em
    T\kern-.1667em\lower.7ex\hbox{E}\kern-.125emX}}

\usepackage{graphicx}
\usepackage{subfigure}
\graphicspath{{figures/}}
\usepackage{amsmath}
\usepackage{amssymb}
\usepackage{xcolor}
\usepackage{siunitx}
\usepackage{amsthm}
\usepackage{multicol}
\usepackage{lipsum}
\usepackage{tikz}

\newcommand{\betamax}{\beta_{\text{max}}}
\newcommand{\alphaB}{\alpha_{\text{B}}}

\newcommand{\betaB}{\beta_{\text{B}}}
\newcommand{\Qmin}{Q_{\text{min}}}
\newcommand{\Qth}{Q_{\text{th}}}
\newcommand{\TB}{T_\text{B}}
\newcommand{\Mt}{M_\text{t}}
\newcommand{\Mr}{M_\text{r}}
\newcommand{\Gt}{G_\text{t}}
\newcommand{\Gr}{G_\text{r}}
\newcommand{\Pt}{P_\text{t}}
\newcommand{\GRIS}{G_\text{RIS}}
\newcommand{\hRIS}{h_\text{RIS}}

\newcommand{\pRIS}{\mathbf{p}_\text{RIS}}
\newcommand{\pSAT}{\mathbf{p}_\text{SAT}}
\newcommand{\pSATu}{\mathbf{p}_{\text{SAT},u}}
\newcommand{\pSATn}{\mathbf{p}_{\text{SAT},n}}
\newcommand{\puRIS}{\mathbf{p}_{u, \text{RIS}}}
\newcommand{\AuRIS}{\mathbf{A}_{u, \text{RIS}}}
\newcommand{\Fun}{F_{u,n}}

\newcommand{\ASATRIS}{\mathbf{A}_\text{SAT,RIS}}
\newcommand{\pSATRIS}{\mathbf{p}_{\text{SAT, RIS}}}
\newcommand{\ru}{r_u}
\newcommand{\su}{s_u}
\newcommand{\vu}{v_u}
\newcommand{\dy}{d_y}
\newcommand{\dz}{d_z}
\newcommand{\pu}{\mathbf{p}_u}
\newcommand{\pun}{\mathbf{p}_{u,n}}

\begin{document}

\title{Enabling NLoS LEO Satellite Communications with Reconfigurable Intelligent Surfaces}

\author{Xiaowen Tian$^{\dag}$, Nuria Gonz\'{a}lez-Prelcic$^{\dag}$, and Takayuki Shimizu$^\ddag$\thanks{ This work has been partially funded by Toyota Motor North America.} \\
	$^{\dag}$ North Carolina State University, Email: \{xtian8,ngprelcic\}@ncsu.edu \\
$^\ddag$ Toyota Motor North America, Email: takayuki.shimizu@toyota.com}

\maketitle

\begin{abstract}
Low Earth Orbit (LEO) satellite communications (SatCom) are considered a promising solution to provide uninterrupted services in cellular networks.
Line-of-sight (LoS) links between the LEO satellites and the ground users are, however, easily blocked in urban scenarios. In this paper, we propose to enable LEO SatCom in non-line-of-sight (NLoS) channels, as those corresponding to links to users in urban canyons, with the aid of reconfigurable intelligent surfaces (RISs). First, we derive the near field signal model for the satellite-RIS-user link. Then, we propose two deployments to improve the coverage of a RIS-aided link: down tilting the RIS located on the top of a building,  and considering a deployment with RISs located on the top of opposite buildings. Simulation results show the effectiveness  of using RISs in LEO SatCom to overcome blockages in urban canyons. Insights about the optimal tilt angle and the coverage extension provided by the deployment of an additional RIS are also provided.
\end{abstract}

%
\IEEEpeerreviewmaketitle

\section{Introduction}
Low Earth Orbit (LEO) satellite communications (SatCom) are becoming a promising solution to the space segment to be integrated with terrestrial networks to provide continuous and global coverage of users \cite{Kodheli2017,TR38.821}. Assuming there is a line-of-sight (LoS) channel between the user and the LEO satellite (SAT), recent work on signal processing for LEO SatCom has focused on the design of beam codebooks \cite{Palacios2022,PalaciosLEO21}, satellite tracking \cite{Khairallah2021}, or massive MIMO designs \cite{Angeletti2020,You2020}. 

Despite the success of LEO SatCom in rural areas and coastal waters, where a LoS link usually exists, its applications in dense urban regions are limited by the  obstructions to the LoS created in complex environments such as urban canyons.  Although there are studies on mitigating blockages in SatCom \cite{Wu2005}, the proposed approaches incur in high latency, which is not tolerable in some use cases such as  vehicular communications. Note that LEO SatCom is also interesting in urban areas to complement terrestrial base stations and provide an alternative connection when a natural disaster like a flooding or a storm causes failure of the terrestrial network.

By creating an additional path to bypass potential blockers, reconfigurable intelligent surfaces (RISs) emerge as a solution for mitigating  blockages in terrestrial communications \cite{DiRenzo2020}. In the context of SatCom, only a few recent works have considered the use of RIS to enhance reception of the SAT signal. Thus, the system model proposed in \cite{GEO} combines a geostationary (GEO) SatCom system with a RIS,
so that the mobile terminal receives the signal through  both the direct link from the SAT and the RIS-assisted link. LEO Satcom aided by RIS has also been proposed in \cite{Matthiesen21} to increase the SNR at the terminal with the same underlying assumption of an existing LoS link with the SAT. Moreover, these works assume that the RIS is operating in the far field mode.

In this paper, we propose the use of RIS to enable LEO SatCom in non-line-of-sight (NLoS) settings, extending the use cases of RIS in SatCom previously proposed. Moreover, unlike prior work of RIS for SatCom, we consider the near-field operation mode for the RIS  \cite{Tang21,Zhang22TWC}, where the relative locations of the RIS elements with respect to the terrestrial terminal need to be considered. This is a practical assumption for our RIS-assisted LEO SatCom system model, given the size of the RIS that needs to be used to provide enough SNR, and the common operation bands for LEO SatCom. We derive the corresponding signal model for the downlink signal received through a large RIS  when the LoS path is blocked, and propose the optimal configuration for the RIS with near field beamforming that maximizes SNR at the SatCom terminal. Simulation results show the effectiveness of different RIS deployments to enable LEO SatCom in a NLoS urban scenario where the buildings in an urban canyon block the direct SAT link.

\section{System Model and Signal Model}
\begin{figure*}
    \centering
    \subfigure[Urban Canyon.]{\includegraphics[width= 3.0 in]{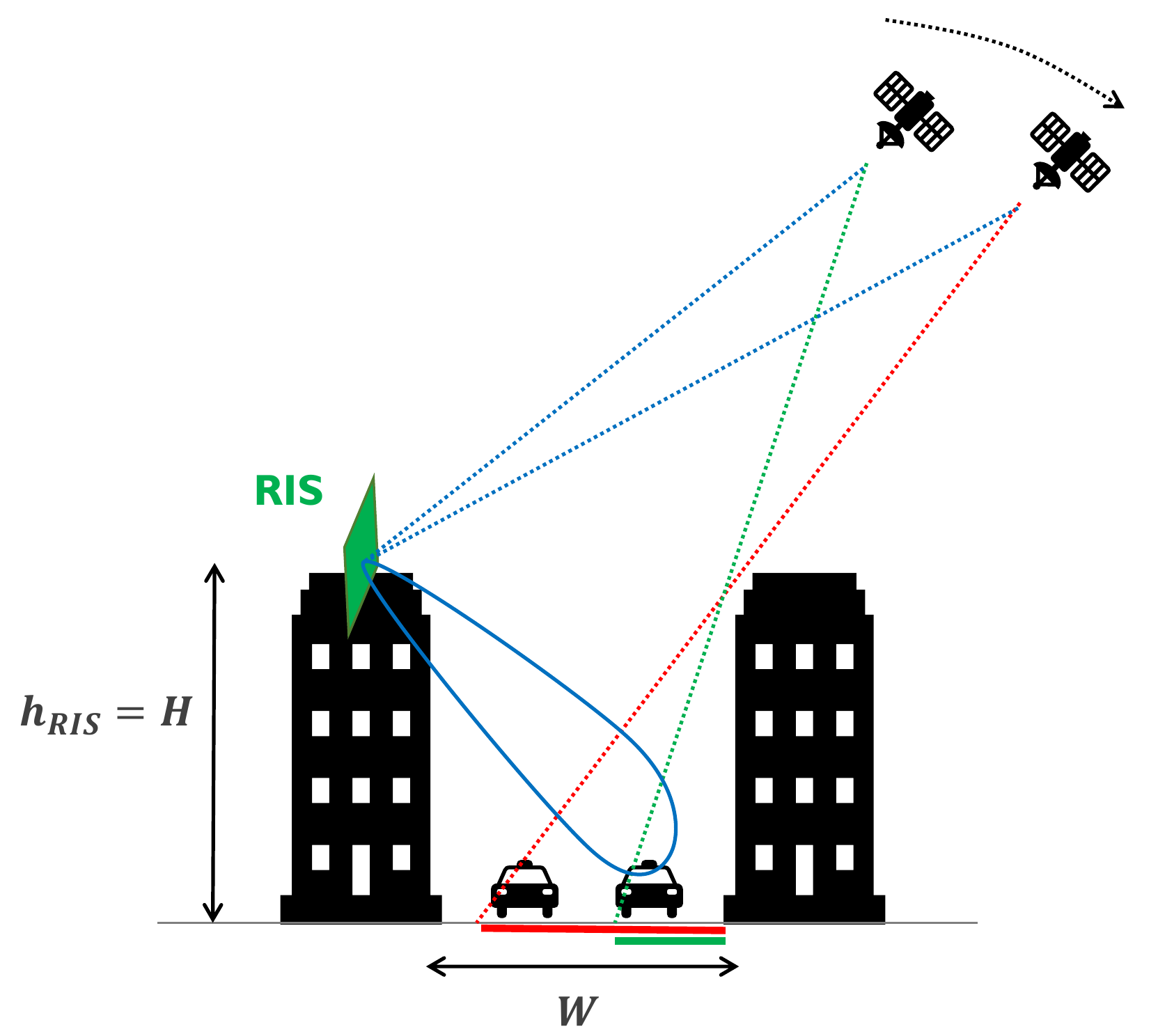}
    \label{fig:urbancanyon}}
    \subfigure[RIS model.]{ \includegraphics[width= 2.5 in]{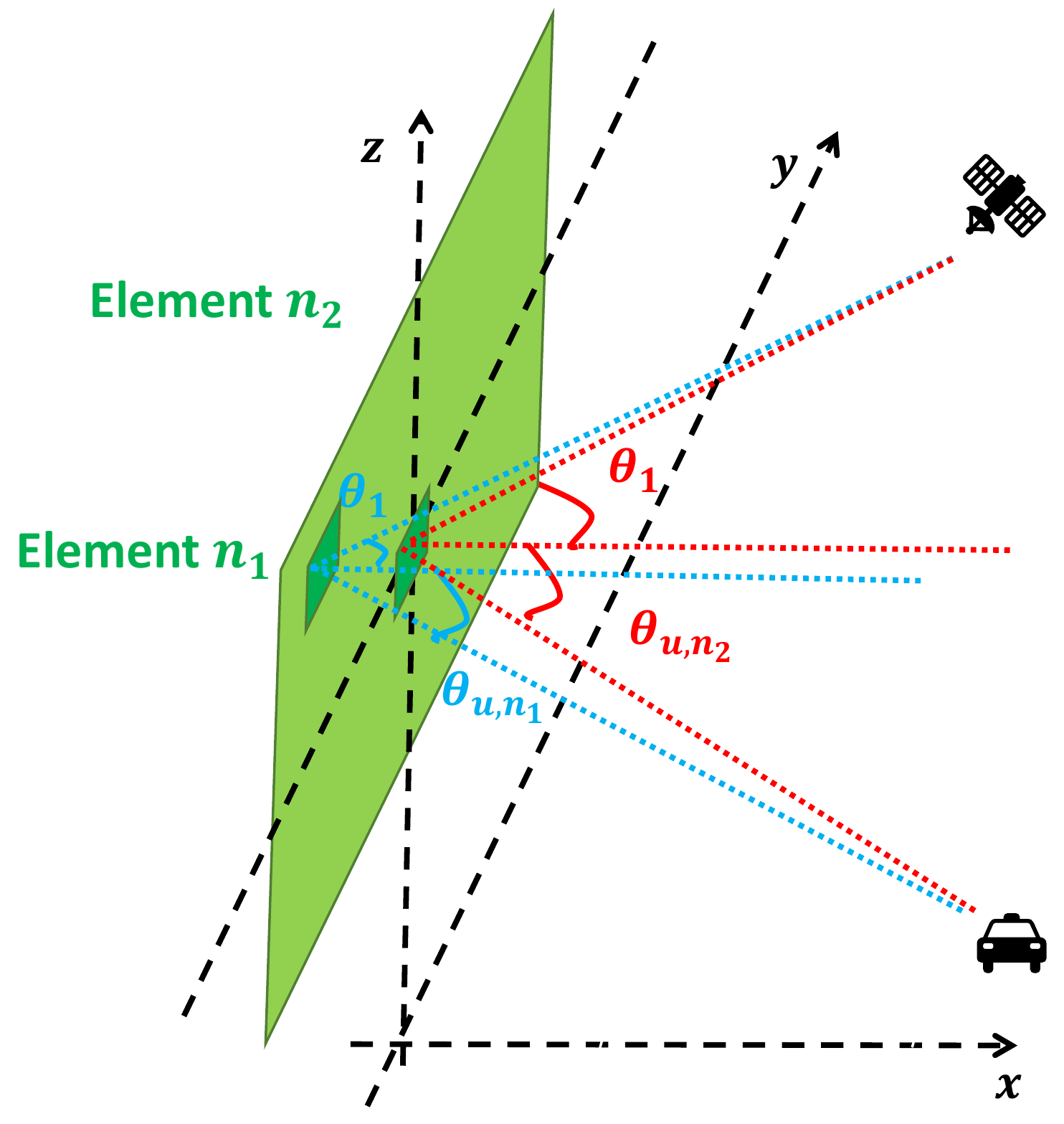}
    \label{fig:RISmodel}}
   \caption{System model. (a) An urban canyon with building height $H$ and street width $W$. The red and green line on the ground illustrates the LoS link blocked region for two different locations of the SAT. A RIS is deployed, with its center located at $(0,0,\hRIS)$ to overcome blockages. The SAT moves along x-axis while the urban canyon is located along y-axis. (b) Elevation angles of a RIS with $N$ elements. $\theta_1$ is the SAT-RIS elevation angle and assumed to be the same since SAT is in the far-field of RIS while $\theta_{u,n}$ is the $n$-the element of RIS to user $u$ link elevation angle which is different for different RIS elements.}
   \label{fig:systemmodel}
\end{figure*}

\subsection{System Model}
We consider a LEO MIMO SatCom system operating in an urban canyon, where buildings are higher than the width of the street, thus blocking the LoS link between the LEO SAT and the user during part of the coverage period, as shown in Fig. \ref{fig:urbancanyon}.
The urban canyon is categorized by the aspect ratio, which is the ratio of canyon height ($H$) to canyon width ($W$), i.e., $\frac{H}{W}$.
Deep canyon is defined as $\frac{H}{W} \approx 2$, regular canyon is $\frac{H}{W} \approx 1$, and avenue canyon is $\frac{H}{W} < 0.5$.
Considering a terminal in an urban canyon, an unblocked direct link with the SAT will exist during a small percentage of the time that the SAT is covering the area, as numerically shown in Section~\ref{sec:whyRIS}. Therefore, to overcome the frequent blockages, we propose to deploy a RIS on the top of the building, i.e.  at a height $\hRIS = H$ and facing the street.

For the ease of derivation, we  only consider a SAT moving from the zenith above the canyon towards the horizon. The initial position of the SAT is denoted as (0, 0, h) while its projection on the ground is the nadir (0, 0, 0).
We also assume that the urban canyon is perpendicular to the movement of the SAT.
The center of the RIS is located at $(0,0, \hRIS)$, as illustrated in Fig.~\ref{fig:RISmodel}. The considered region of interest in the urban canyon has a width $W$ and a length $L$.

We assume that the SAT is equipped with $\Mt$ antennas and the user has $\Mr$ antennas.
The RIS has $N = N_\text{y} \times N_\text{z}$ elements distanced by half-wavelength, placed vertical to the ground.
Because of the large distance between the SAT and the RIS, a high number of elements is used.
However, the increase of the RIS aperture will also increase its near-far field boundary, which can be characterized by the Fraunhofer distance as $\frac{2D^2}{\lambda}$ \cite{Tang21}.
It will be shown in Section~\ref{sec:results} that a RIS supporting a typical LEO SatCom link will be operating in the near-field.
Therefore, we need to derive a a near-field signal model to describe the received signal for the SAT-RIS-user link.

\subsection{Near-field RIS-assisted link signal model}
The RIS is modeled as a network of $N$ phase shifters, each one of them with a phase configuration $\psi_n$, $n=1,\ldots, N$. The RIS configuration matrix is defined as  $\Omega$, satisfying
\begin{equation}
\mathrm{diag}(\mathbf{\Omega}) = [ e^{j \psi_1}, ..., e^{j \psi_N} ]  = \boldsymbol{\psi}.
\end{equation}

We use the locations of each RIS element illustrated in Fig.~\ref{fig:RISmodel} to develop the near-field signal model. The location of the center of the RIS is $\pRIS = (0,0,H)$, and the location of the SAT is $\pSAT$.
The location of the $n$-th element of RIS is $\mathbf{p}_{n}$ and the location of the user is $\mathbf{p}_{u}$.
Then the link distance between the SAT and the $n$-th RIS element is
$|\pSATn|$, with $\pSATn = \pSAT - \mathbf{p}_{n}$.
Similarly, the link distance between user and the $n$-th element is $|\pun|$, with $\pun = \pu - \mathbf{p}_{n}$.

We describe now the near-field MIMO channel between the RIS and the user as steering vectors combined with an attenuation term.
Given the direction of the user, since the user's antenna array works in the far-field mode, we can use the far field steering vector $\mathbf{c}(\pu)$ to describe the impact of the user array in the MIMO channel matrix. For the steering vector of the RIS, we consider the definition \cite{Zhang22TWC}
\begin{equation}
    \mathbf{a}(\puRIS) = [e^{-j k |\mathbf{p}_{u,1}|},  ... , e^{-j k |\mathbf{p}_{u, N}|}]^T,
    \label{eq:au}
\end{equation}
with $k=\frac{2 \pi}{\lambda}$ being the wave number.
Let us denote the channel RIS-user as $\mathbf{H} \in \mathbb{R}^{\Mr \times N}$. It can be written as
\begin{equation}
     \mathbf{H} =  \mathbf{c}(\pu)  \mathbf{a}^T(\puRIS) \AuRIS,
     \label{eq:h}
\end{equation}
where the attenuation term $\AuRIS$ is a diagonal matrix with 
\begin{equation}
    \mathrm{diag} (\AuRIS) = \left[\frac{\sqrt{F_{u,1}}}{|\mathbf{p}_{u,1}|}, ..., \frac{\sqrt{F_{u,N}}}{|\mathbf{p}_{u, N}|} \right].
    \label{eq:Au}
\end{equation}
The radiation profile for each RIS element is  \cite{Tang21}
\begin{equation}
    F(\theta_n, \phi_n) = \left\{
    \begin{array}{ll}
          \cos^b\theta_n, & \theta_n \in [0, \frac{\pi}{2}], \\
          0, & \mathrm{otherwise}.
    \end{array}
         \right.
         \label{eq:F}
\end{equation}
In (\ref{eq:F}), $b$ determines the bore-sight gain, and $\theta_n$ and $\phi_n$ denote the elevation and azimuth angles between the $n$-th element of the RIS and the user.
The gain of the $n$-th element of the RIS is then determined by $F(\theta_n, \phi_u)$ as \cite{Tang21}
\begin{equation}
    \GRIS = \frac{4\pi}{\int_{\phi_n=0}^{2\pi} \int_{\theta_n=0}^{\pi} F(\theta_n, \phi_n) \sin\theta_n d\theta_n d\phi_n}.
\end{equation}

Let us denote the channel SAT-RIS as $\mathbf{G} \in \mathbb{R}^{N\times \Mt}$, that can be written as
\begin{equation}
    \mathbf{G} = \ASATRIS \mathbf{a}(\pSATRIS) \mathbf{b}^H(\pSAT),
    \label{eq:G}
\end{equation}
where the steering vector for the RIS
is defined as
\begin{equation}
    \mathbf{a}(\pSATRIS) = [e^{-j k |\mathbf{p}_\text{SAT,1}|},  ... , e^{-j k |\mathbf{p}_{\text{SAT},N}|}]^T,
    \label{eq:aSAT}
\end{equation}
while the steering vector at the SAT side is $\mathbf{b}(\pSAT)$, which follows the far field model since the SAT's antenna works in the far-field mode. The attenuation matrix for this channel is also a diagonal matrix with elements
\begin{equation}
    \mathrm{diag} (\ASATRIS) = \left[\frac{\sqrt{F_\text{SAT,1}}}{|\mathbf{p}_\text{SAT,1}|}, ..., \frac{\sqrt{F_{\text{SAT},N}}}{|\mathbf{p}_{\text{SAT},N}|} \right].
    \label{eq:ASAT}
\end{equation}

The received signal at user $u$ is
\begin{equation}
    \ru = \mathrm{coef}  \mathbf{w}^H \mathbf{H} \mathbf{\Omega} \mathbf{G} \mathbf{f} \su + \vu,
    \label{eq:ru}
\end{equation}
where $\vu$ is the noise with covariance $\sigma^2$, and $\mathrm{coef}$ is a fixed coefficient irrelevant for the phase shifters configuration and deployment of the RIS. It can be written as
\begin{equation}
    \mathrm{coef} = \sqrt{ \Pt \Gt \Gr \GRIS \dy \dz } \frac{\lambda}{4 \pi},
    \label{eq:coef}
\end{equation}
where $\Pt$ is the transmit power, $\Gt\Gr$ is the combined antenna gain created by the transmitter and receiver, $\dy\dz$ is the size of one RIS element, and $\GRIS$ is the gain for each RIS element.

In this paper, we assume that the beamformers at both the SAT and users are optimal, i.e.,  $\mathbf{f} = \mathbf{b}(\pSAT)$ and $\mathbf{w} = \mathbf{c}(\pu)$.
Under the far-field assumption, we can user the far-field RIS parameters to replace the near-field RIS parameters, i.e., $F_{1}(\theta_1) = F_{\text{SAT},n}(\theta_{\text{SAT},n})$ and $ \mathbf{p}_{1} = \pSAT-\pRIS$,
where $F_{1}(\theta_1)$ denotes the radiation profile of the RIS's center to the SAT with elevation angle $\theta_1$. After defining the x-axis $\mathbf{u}_x(1,0,0)$, the elevation angles can be obtained as
\begin{equation}
    \cos\theta_1 = \frac{|<\mathbf{p}_{1}, \mathbf{u}_x>|}{|\mathbf{p}_{1}|}, \cos\theta_{u,n} =  \frac{|<\pun, \mathbf{u}_x>|}{|\pun|}.
\end{equation}
Then, the received signal $ \ru$ can be written as (\ref{eq:rufinal}), at the top of next page.
\begin{figure*}
\begin{equation}
    \begin{aligned}
    \ru &=  \mathrm{coef} \sum^{N}_{n=1} e^{-jk |\pun|}  \frac{\sqrt{\Fun(\theta_{u,n})}}{|\pun|} e^{j \psi_n} \frac{\sqrt{F_{\text{SAT},n}(\theta_{\text{SAT},n})}}{|\pSATn|} e^{-jk |\pSATn|} \su + \vu \\
    &=  \mathrm{coef} \frac{\sqrt{F_1(\theta_1)}}{|\mathbf{p}_{1}|} \sum^{N}_{n=1} \frac{\sqrt{\Fun(\theta_{u,n})}}{|\pun|} e^{j \psi_n} e^{-jk (|\pun| + |\pSATRIS|)} \su + \vu,
\end{aligned}
\label{eq:rufinal}
\end{equation}
\end{figure*}
The optimal configuration for the $n$-th RIS element can be obtained by solving the next optimization problem
\begin{equation}
    \boldsymbol{\psi} = \arg \max \left| \boldsymbol{\psi}^T \left(  \mathbf{a}\left(\puRIS\right) \circ \mathbf{a}\left(\pSATRIS\right)  \right) \right|^2,
\end{equation}
where $\circ$ represents the element-wise product,
with a closed form solution
\begin{equation}
    \psi_n^*  = \mathrm{mod} \left( \frac{2 \pi}{\lambda} \left( \left|\pun\right| + \left|\mathbf{p}_{1}\right| \right),  2\pi \right).
    \label{eq:RISconf}
\end{equation}
When the beamformers and the RIS are both optimally configured, the received signal becomes
\begin{equation}
    \ru = \mathrm{coef} \frac{\sqrt{F_1(\theta_1)}}{|\mathbf{p}_{1}|} \sum^{N}_{n=1} \frac{\sqrt{\Fun(\theta_{u,n})}}{|\pun|}  \su + \vu.
\label{eq:ruF}
\end{equation}
Finally, the SNR of the SAT-RIS-user link can be written as
\begin{equation}
    \gamma_\text{RIS} =  \frac{ \Pt \Gt \Gr \GRIS \dy \dz}{\sigma^2} (\frac{\lambda}{4 \pi})^2 \frac {F_1(\theta_1)} {|\mathbf{p}_{1}|^2}   \left| \sum^{N}_{n=1} \frac{ \sqrt{ \Fun(\theta_{u,n}) } }{|\pun|} \right|^2.
    \label{eq:SNRRIS}
\end{equation}

\subsection{LoS signal model}
The signal received through the LoS link follows a far field model, and can be written as
\begin{equation}
    y_u = \sqrt{\Pt \Gt \Gr} \frac{\lambda}{4 \pi | \pSATu|} \mathbf{b}^H( \pSATu ) \mathbf{f} \su + \vu,
    \label{eq:yu}
\end{equation}
where $\mathbf{f} = \mathbf{b}( \pSATu )$ to make sure that the SAT is always pointing to the location of user $u$.
From (\ref{eq:yu}), with optimal beamforming, the received signal strength of the LoS link is only affected by the distance of SAT and user $u$, i.e., $| \pSATu|$.
Then the SNR of the LoS link is
\begin{equation}
    \gamma_\text{LoS} = \frac{\Pt \Gt \Gr}{\sigma^2} (\frac{\lambda}{4 \pi})^2 \frac{1}{ |\pSATu|^2 }.
    \label{eq:SNRLoS}
\end{equation}

For comparison purposes, in this paper we assume that the user can only be served by either the LoS link or the RIS-assisted link.

\section{Enabling NLoS SatCom with RIS}

In this section we will first justify the need of a technology capable to increase coverage of LEO SatCom in urban scenarios. Then we will propose two different RIS deployments to enhance coverage by using RIS.

\subsection{Why RIS for LEO SatCom?}\label{sec:whyRIS}
\begin{figure}
    \centering
    \begin{tabular}{cc}
    \includegraphics[width = 2.3 in]{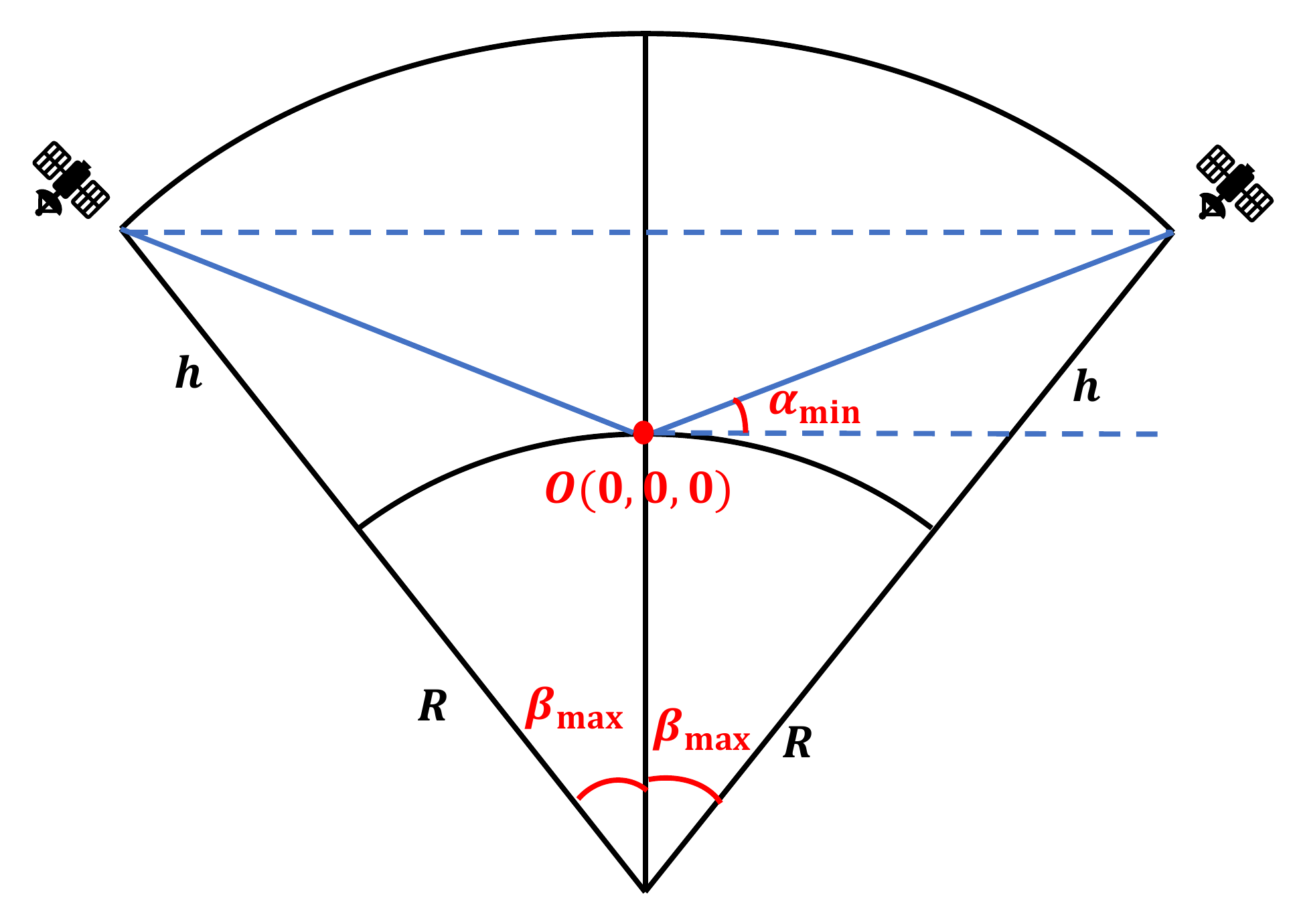}&
   \hspace*{-7mm} \includegraphics[width = 1.2 in]{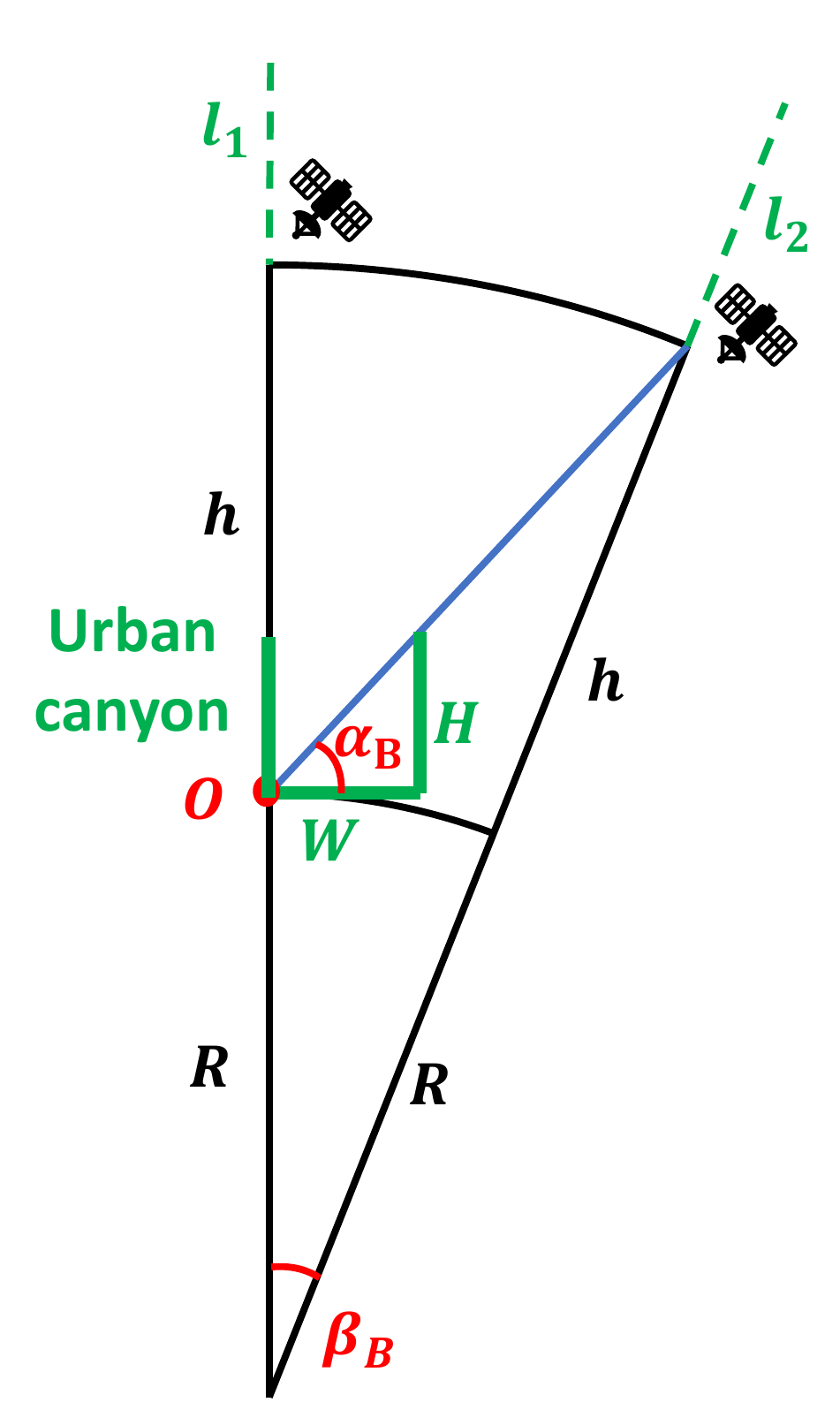}
   \\
   (a) & (b)
    \end{tabular}
    \caption{Illustration of coverage and blockage  of the LEO SatCom direct link in an urban canyon: (a) geometric relationships; (b) beginning of blockage.}
    \label{fig:blockage}
\end{figure}

LEO SatCom deployments cannot provide in general a continuous coverage in urban environments. In these settings, when the channel is NLoS because the buildings block the LoS component between the user and the SAT, it is not possible to create a sufficiently strong link to enable high data rate transmission, even if the channel is perfectly known. 
Furthermore, it is not only that occasionally the SAT link becomes NLoS. In general, when considering an urban setting, the canyons created by the buildings will  block the LoS path a high percentage of the time a given SAT is orbiting over a city. To be more specific, we compute the percentage of time the LEO SatCom link is blocked, denoted as $\TB$, assuming different types of urban canyons. Assuming that the orbit of the SAT passes right above  the urban canyon, this value depends on the SAT height and the number of SATs per orbit, denoted as $Q$. To obtain the expression for $\TB$, we define first $2\betamax$ as the angular region covered by one LEO SAT, which is obtained as $\betamax = \frac{\pi}{Q}$, as illustrated in Fig.~\ref{fig:blockage}(a). The SAT elevation angle $\alphaB$, where the blockage created by a given canyon of height $H$ starts, fulfills  $\tan\alphaB = \frac{H}{W}$. This parameter has a geometric relationship with the angle  $\betaB$, as illustrated in Fig.~\ref{fig:blockage}(b).
With these definitions, the blockage ratio can be obtained as
\begin{equation}
    \TB = 1- \frac{\betaB}{2\betamax}.
\end{equation}
Table~\ref{tab:selectedconst} shows some examples of the values that  $ \TB$ takes for some commercial constellations. The LoS link is blocked most of the time that a LEO SAT is covering an urban area.
Generally, this is because in commercial constellations it is not possible to deploy as many SATs as needed to guarantee a LoS link all the time. The minimum number of SATs to guarantee LoS coverage can be computed as
\begin{equation}
    \Qmin = \frac{2\pi}{\betaB}.
    \label{eq:Qmin}
\end{equation}
This value is always higher than the number of SATs practically deployed. For example, in Starlink Phase 1, Shell 1,  the number of SATs per orbit is 22, while for a canyon with ratio $\frac{H}{W}=1.4$, $\Qmin=113$.
The need for a technology that can extend LEO SatCom coverage in urban scenarios is obvious.

\begin{table}
\caption{Blockage ratio in urban canyons  for selected constellations.}
\label{tab:selectedconst}
\centering
\begin{tabular}{ c||c|c }
\hline
Constellation &  Canyon $\frac{H}{W}$ & Blockage ratio $\TB$ \\
\hline
\hline
\multirow{3}{12 em}{Telesat, polar orbit} &  1.4 & $80.3\%$  \\
 & 2 & $86.0\%$  \\
& 2.4 & $88.3\%$  \\
\hline
\multirow{3}{12 em}{Telesat, inclined orbit}  & 1.4 & $79.3\%$  \\
 & 2 & $85.2\%$  \\
& 2.4 & $87.6\%$ \\
\hline
\multirow{3}{12 em}{Starlink Phase 1, Shell 1} & 1.4 & $80.5\%$  \\
& 2 & $86.2\%$  \\
 & 2.4 & $88.5\%$  \\
\hline
\multirow{3}{12 em}{Starlink Phase 1, Shell 2} &  1.4 & $81.7\%$ \\
 & 2 & $87.1\%$  \\
& 2.4 & $89.2\%$  \\
\hline
\multirow{3}{12 em}{Starlink Phase 1, Shell 3} &  1.4 & $47.8\%$ \\
 & 2 & $63.1\%$ \\
& 2.4 & $69.1\%$  \\
\hline
\end{tabular}
\end{table}

\subsection{Tilted RIS-Assisted LEO SatCom}
From Fig. \ref{fig:systemmodel}, it is obvious that the LoS link is likely to be blocked with smaller SAT elevation angles $\theta_1$, that appear when the SAT is moving further away. But at the same time, the SNR for the RIS link will be larger since the term $\cos\theta_1$ in (\ref{eq:SNRRIS}) will increase with a decreasing SAT elevation angle $\theta_1$. This means that when the SAT is above the user, the LoS link will serve the users, and when the
 the SAT is moving away, the LoS link tends to be blocked, but the additional link supported by the RIS can overcome the blockage and serve the user.
However, according to (\ref{eq:SNRRIS}), the SNR of the RIS link for near users, i.e., users with large elevation angle $\theta_{u,n}$ because they are close to the building where the RIS is deployed, will be low due to a smaller $\cos\theta_{u,n}$. This makes the RIS coverage for near users a challenge.

We can exploit the same idea used in cellular communications of down tilting the antennas at the base stations to get a higher SNR for near users. Thus, with the SAT being in the far-field of the RIS, after down tilting the RIS by an angle $\theta_0$, the distance between the SAT and the center of the RIS $\mathbf{p}_{1}$ will remain the same, while the new elevation angle becomes $\Tilde{\theta}_1 = \theta_1 + \theta_0$. Because the  users are in the near-field of the RIS, after tilting the RIS, the location of the $n$-th element $\mathbf{p}_{n}$ also changes.

Let us denote the location of the $n$-th element before tilting as $\mathbf{p}_{n} = (0, y_n, z_n+H)$.
After down tilting the RIS, the location of the $n$-th element is $\Tilde{\mathbf{p}}_{n} = (z_n\sin\theta_0, y_n, z_n\cos\theta_0+H)$. To find the new elevation angle $\Tilde{\theta}_{u,n}$, we should compute the angle between the vector $\Tilde{\mathbf{p}}_{u,n}$ and $\Tilde{\mathbf{u}}_x(\cos\theta_0, 0, \sin\theta_0)$. Thus,  $\cos\Tilde{\theta}_{u,n}$ can be written as
\begin{equation}
    \cos\Tilde{\theta}_{u,n} = \frac{| (x_u - z_u \sin\theta_0)\cos\theta_0 - (z_n\cos\theta_0 + H)\sin\theta_0 |}{|  \Tilde{\mathbf{p}}_{u,n} |}.
\end{equation}
The SNR of the SAT-RIS-user link with a down tilted RIS is then given as
\begin{equation}
    \Tilde{\gamma}_\text{RIS} =  \frac{ \Pt \Gt \Gr \GRIS \dy \dz}{\sigma^2} (\frac{\lambda}{4 \pi})^2 \frac {F_1(\Tilde{\theta}_1)} {|\mathbf{p}_{1}|^2}   \left| \sum^{N}_{n=1} \frac{ \sqrt{ \Fun(\Tilde{\theta}_{u,n}) } }{|\Tilde{\mathbf{p}}_{u,n}|} \right|^2.
    \label{eq:SNRRIS_tilt}
\end{equation}

The optimal tilt angle $\theta_0^*$ for different users and SAT locations is different, and it can be obtained by solving for the maximum $\Tilde{\gamma}_\text{RIS}^*$ as in (\ref{eq:SNRRIS_tilt}).
We will provide some insights about the optimal tilting angle $\theta_0^*$ in Section~\ref{sec:results} .

\subsection{Double RIS-Assisted SatCom}

\begin{figure}
    \centering
    \includegraphics[width = 3.0in]{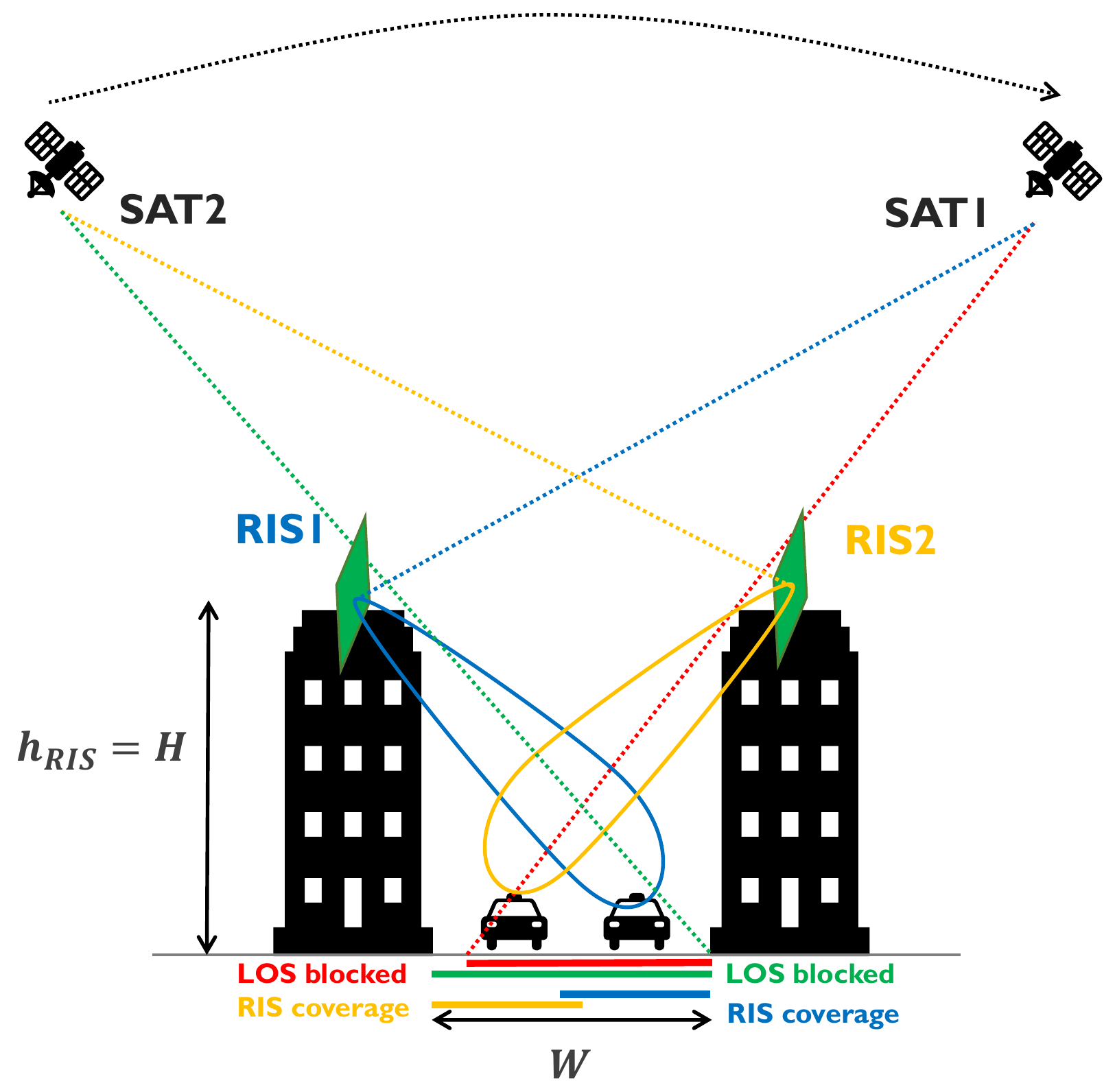}
    \caption{Illustration of blockages under urban canyon with two SATs and two opposite deployed RISs.}
    \label{fig:TwoRISBlockage}
\end{figure}

Another way to improve coverage is increasing the number of RISs serving a given region.
Thus, we assume now that two RISs are deployed facing each other in opposite buildings in a street, and two SATs orbiting at both sides of the canyon have LoS links to those two RISs, as illustrated in  Fig. \ref{fig:TwoRISBlockage}.

We denote the SAT at the right side as  SAT1, which has a LoS link with RIS1 on the left building.
Similarly, the SAT at left side of the canyon is SAT2, which has a LoS link with RIS2, on the right building.
Since the SATs are far from the RISs and the user located at $O$, we approximate the elevation angle of the SAT-RIS links as the elevation angle of user $O$, i.e., $\theta_s = \alpha_s$, with $s$ representing the index of SAT/RIS.

Let us define a threshold value for the number of SATs per orbit  $\Qth$, as the maximum number of SATs in the constellation so only one SAT at a time can be visible assuming from the ground with no obstructions. Graphically, this situation appears when the straight line  connecting both SATs in Fig.~\ref{fig:blockage}(a) passes though the origin $O$.  According to the number of SAT per orbit $Q$, there are three cases regarding the LoS connection of ground users and the number of serving SATs.

\begin{itemize}
    \item $Q < \Qth$, with the threshold of number of SAT per orbit denoted as $\Qth$.
    In this case, there's only one SAT above the horizon.
    Despite there are multiple SATs in one orbit, the serving status is the same as with a single RIS case, because only one RIS is illuminated by the SAT. For a number of SATs per orbit $\Qth$, the SAT angle separation is set to be $\beta_{\Qth}$, fulfilling
    \begin{equation}
    \cos{\frac{\beta_{\Qth}}{2}} = \frac{R}{R+h}, \Qth = \lfloor \frac{2\pi}{\beta_{\Qth}} \rfloor,
    \label{eq:Qth}
    \end{equation}
    where $R$ is the Earth's radius and $\lfloor . \rfloor$ is the floor function.
\item  $\Qth < Q < \Qmin$, where $\Qmin$ is determined only by the urban canyon parameters as in (\ref{eq:Qmin}).
In this case, the ground users will be partially blocked during the movement of the SATs.
\item $Q > \Qmin$, LoS links between the SAT and the ground users will always be available, and applying RISs will be useful to increase the SNR only, since there are no blockages caused by the urban canyon.
\end{itemize}

In this paper we focus on the second case,  i.e. $\Qth < Q < \Qmin$, and provide numerical results to illustrate the significant coverage  improvement that can be achieved by leveraging two RISs deployed on opposite buildings in the street.
%

\section{Simulation Results}\label{sec:results}

\begin{figure*}[t]
  \centering
 \subfigure{\includegraphics[width= 2.1 in]{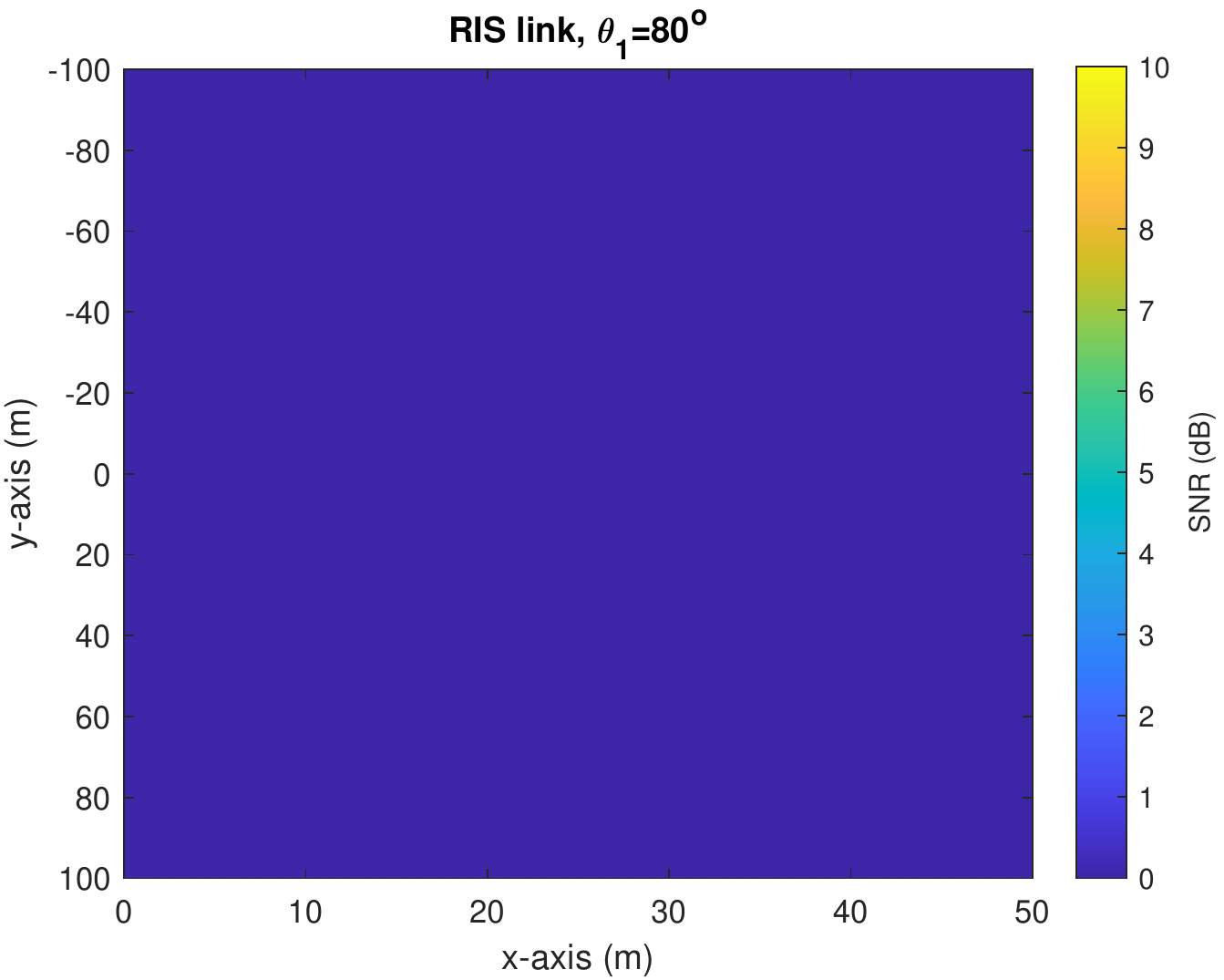}}
 \hspace{0em}
 \subfigure{\includegraphics[width= 2.1 in]{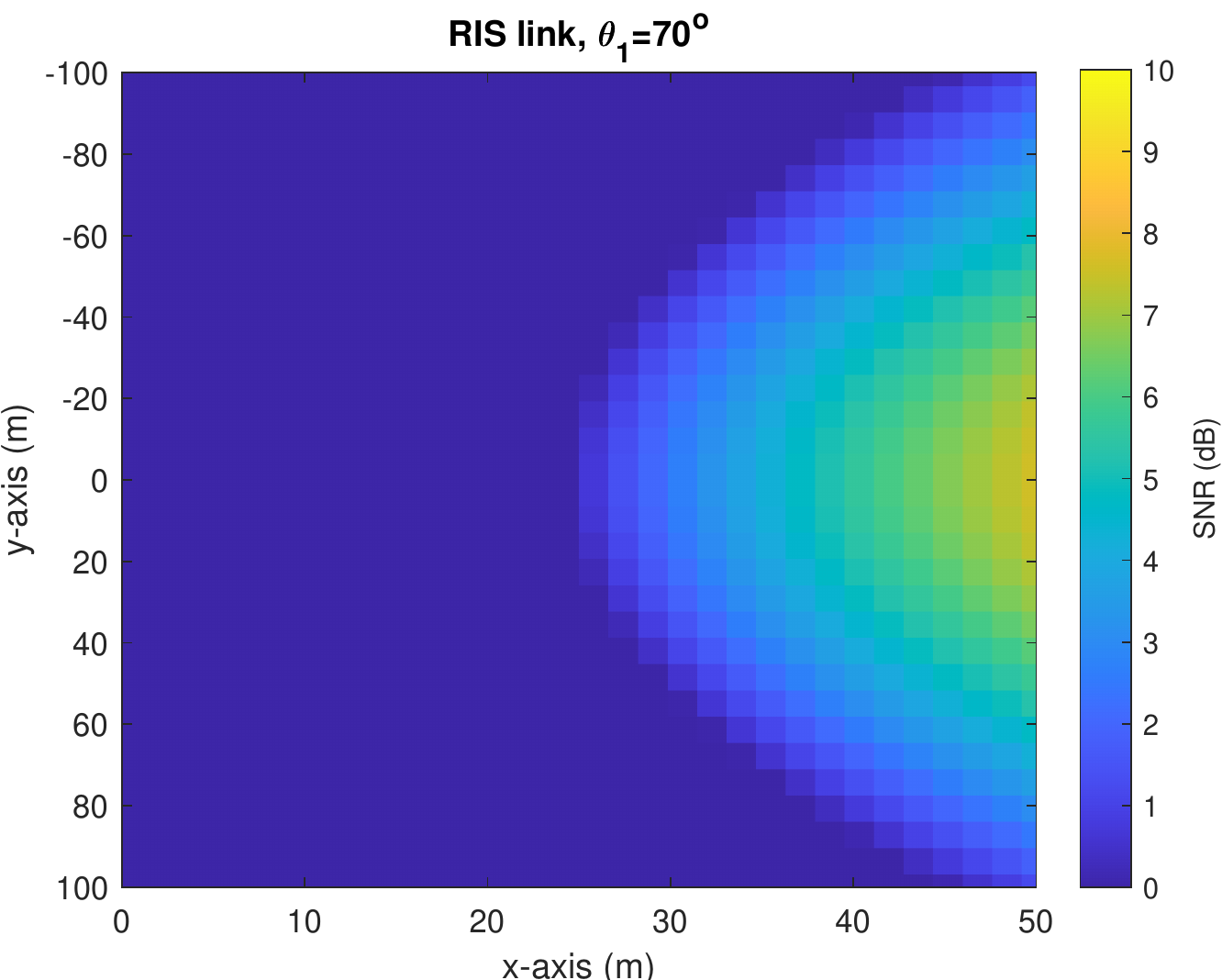}}
 \hspace{0em}
 \subfigure{\includegraphics[width= 2.1 in]{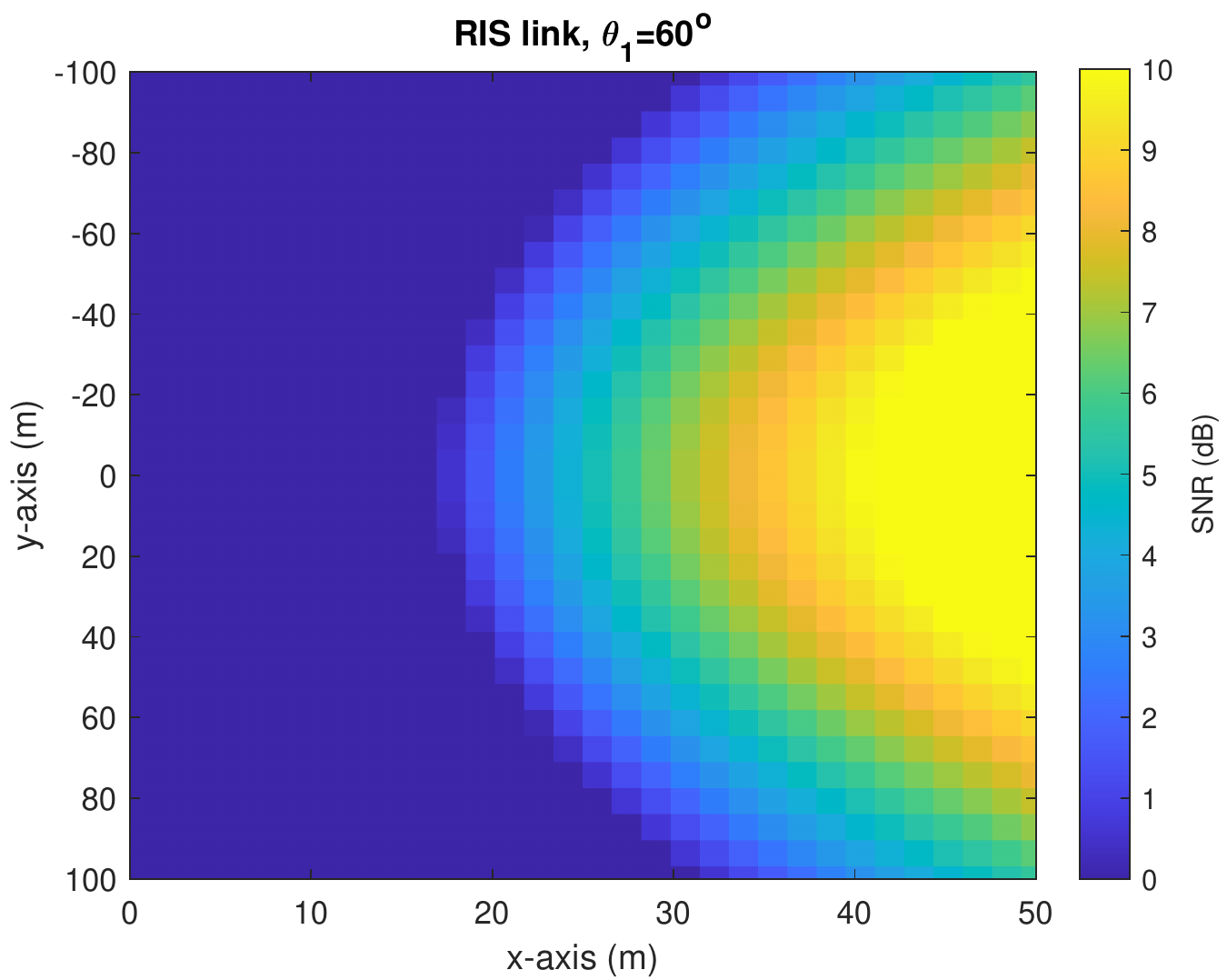}}
 \hspace{0em}
   \vspace{-0.0cm}
  \caption{Coverage of the SAT-RIS link for different SAT elevation angles.}
  \label{fig:Comparison1}
  \vspace{-0.0cm}
\end{figure*}


We simulate the downlink of a RIS-aided LEO SatCom system operating at $11.54$ GHz, with the SAT height $h = 1300$ km.
The transmit power is set as $\Pt = 15$ dBW, and the atmospheric path loss is $0.0166$ dB.
The transmitter gain is set to $\Gt = 24.6$ dB, that corresponds to employing $12\times24$ subarrays to create the different beams illuminating a given region. The receiver gain is set to $\Gr = 27.6$ dB, which corresponds to a $24\times24$ planar array at the user side.
The noise power is $\sigma^2 = -120.5$ dBW, corresponding to a noise temperature of $24.1$ dBK and a bandwidth of $250$ GHz.
The RISs are of size $5$ m length and $3$ m width, and the distances between each element is half-wavelength. The urban canyon dimensions are $H=100$ m and $W=50$ m.

\subsection{Coverage of a single RIS}
\label{sec:singleRIS}

The coverage heat maps of a single RIS are shown in Figs.~\ref{fig:Comparison1} for different elevation angles of the satellite, when the RIS is located at $(0,0)$, the center of the left boundary of the considered urban canyon region. 
The shape of the map is the result of the elevation term in the SNR expression (\ref{eq:SNRRIS}), where the users at the right side will have smaller elevation angles $\theta_{u,n}$. Note that a single RIS will not cover the users at the left side.
Furthermore, with large SAT elevation angle such as $\theta_1=80^\circ$, the RIS-assisted link does not provide coverage.

%
\begin{figure}
    \centering
    \includegraphics[width = 2.9 in]{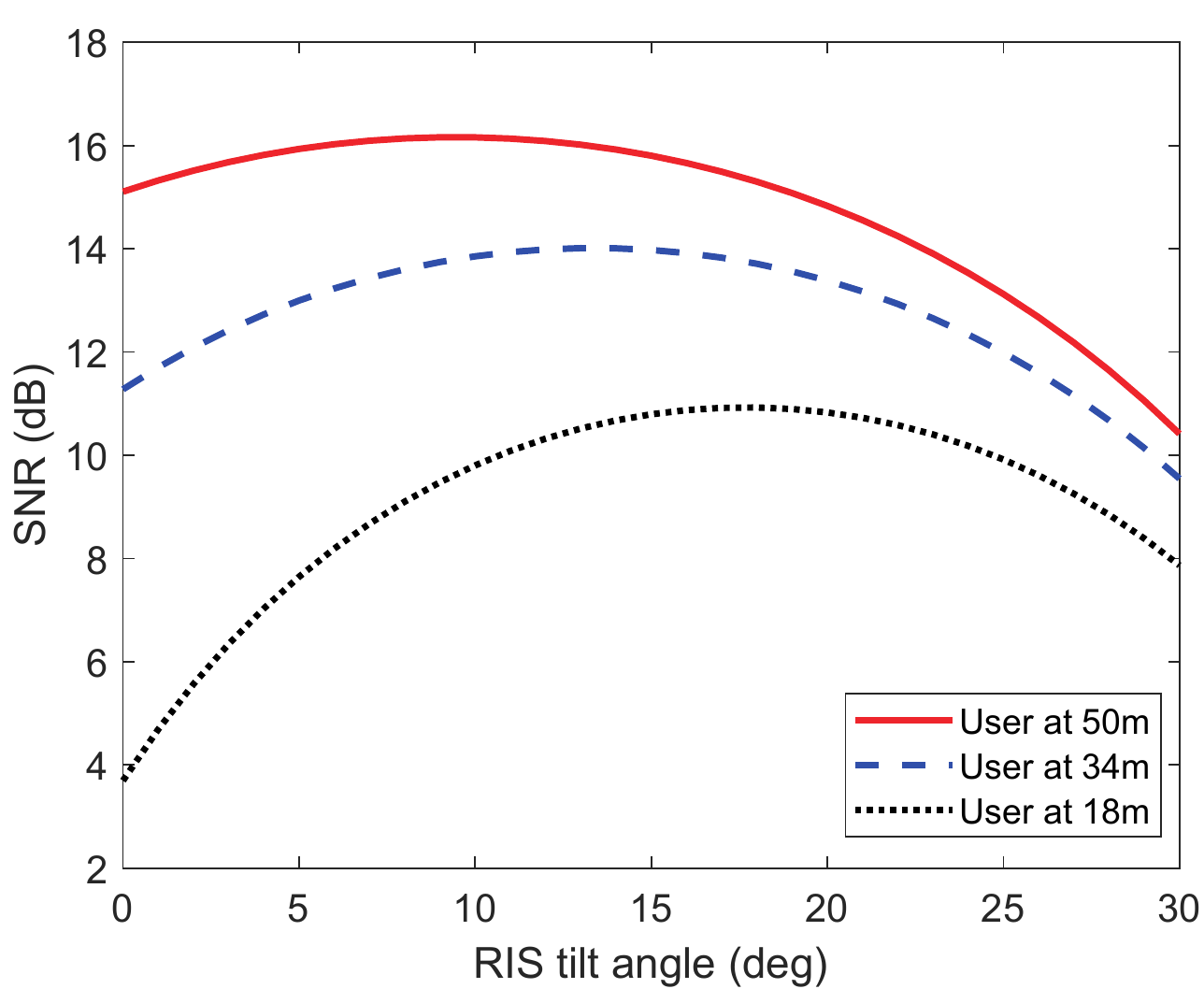}
    \caption{SNR versus different RIS tilt angle under deep canyon with SAT location of $\theta_1=45^o$.}
    \label{fig:UserLocation}
\end{figure}
\subsection{Tilted RIS-Assisted SatCom}
From Figs.~\ref{fig:Comparison1}, it is straightforward to conclude that a RIS deployed vertically on the top of a building cannot serve the users on the ground close to that same building.
We evaluate now the coverage improvement that can be achieved by down tilting the RIS by an angle $\theta_0$.
In Fig. \ref{fig:UserLocation}, we show the SNR versus the RIS tilt angle for users located at different points.
It can be seen that the performance of near users can be severely decreased with non-optimal tilt angle, while the performance of far users remains satisfactory.

\subsection{Double RIS-assisted SatCom}
\begin{figure}[t]
  \centering
  \subfigure{\includegraphics[width= 2.9 in]{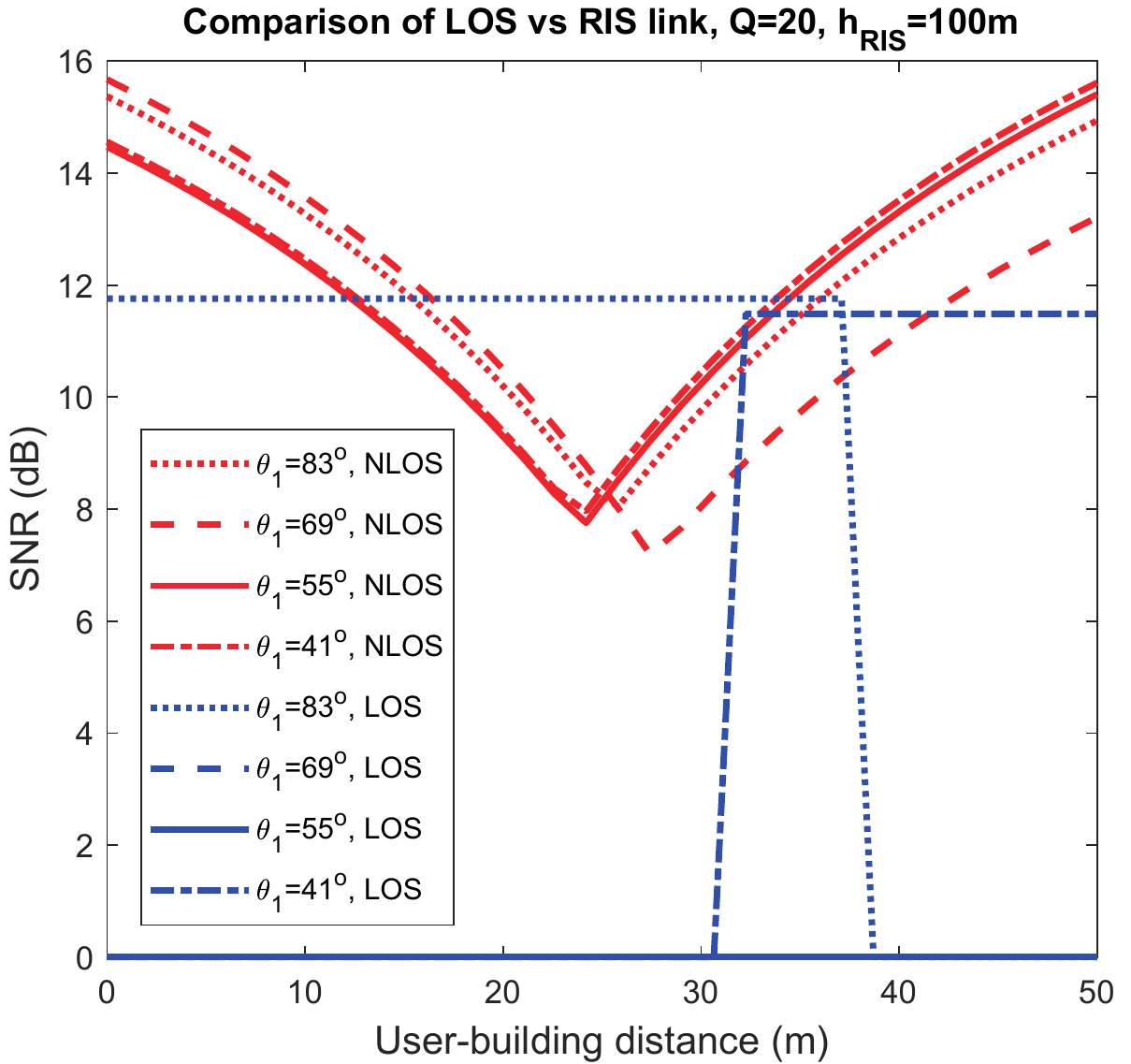}}
  \hspace{0em}
   \vspace{-0.0cm}
  \caption{Comparison between LoS links and RIS-assisted links. Received SNR versus user-building distance under different SAT1 elevation angle.}
  \label{fig:OppoUserdistance}
  \vspace{-0.0cm}
\end{figure}


With $h= 1300$ km, it can be calculated that $\Qmin = 75$ and $\Qth=5$.
To have at least two SATs above the horizon, and focusing on the cases where the LoS link can be blocked, we study the scenario where $5<Q<75$.
To ensure a satisfactory RIS-assisted link coverage we consider elevation angles for the SAT smaller than $65^\circ$. More complex deployments of SATs and RIS will be needed to guarantee full coverage.
The received SNR versus user-building distance under different SAT1 elevation angle and for $Q=20$ are shown in Fig. \ref{fig:OppoUserdistance}.
Comparing each pair of LoS and RIS links (red lines with blue lines), under our RIS-aided strategy, RIS-assisted link overcomes the blockages in the deep canyon scenario for all SAT elevation angles while the LoS link can be partially blocked.
With $Q$ increasing, the LoS blockage ratio will be decreasing, and the RIS links always have satisfactory SNRs.

\section{Conclusions}
In this paper, we proposed to enable NLoS  LEO SatCom with RISs, overcoming the frequent blockages that appear in urban scenarios. We considered a realistic signal model that assumes operation in the near field, unlike most of the work on RIS-aided communication. We proposed two possible RIS deployments to improve coverage,  and their performance was verified through simulation studies.
Down tilting the RIS was shown to improve near user coverage with small SAT elevation angles. Deploying two opposite RISs was shown to provide full coverage for all users when more than one SAT is available to illuminate the scenario.

\bibliographystyle{IEEEtran}
\bibliography{Xiaowen}

\begin{thebibliography}{10}
\providecommand{\url}[1]{#1}
\csname url@samestyle\endcsname
\providecommand{\newblock}{\relax}
\providecommand{\bibinfo}[2]{#2}
\providecommand{\BIBentrySTDinterwordspacing}{\spaceskip=0pt\relax}
\providecommand{\BIBentryALTinterwordstretchfactor}{4}
\providecommand{\BIBentryALTinterwordspacing}{\spaceskip=\fontdimen2\font plus
\BIBentryALTinterwordstretchfactor\fontdimen3\font minus
  \fontdimen4\font\relax}
\providecommand{\BIBforeignlanguage}[2]{{%
\expandafter\ifx\csname l@#1\endcsname\relax
\typeout{** WARNING: IEEEtran.bst: No hyphenation pattern has been}%
\typeout{** loaded for the language `#1'. Using the pattern for}%
\typeout{** the default language instead.}%
\else
\language=\csname l@#1\endcsname
\fi
#2}}
\providecommand{\BIBdecl}{\relax}
\BIBdecl

\bibitem{Kodheli2017}
O.~{Kodheli} \emph{et~al.}, ``{Integration of Satellites in 5G through LEO
  Constellations},'' in \emph{Proc. of GLOBECOM}, 2017, pp. 1--6.

\bibitem{TR38.821}
{Technical Specification Group Radio Access Network}, ``{{TR38.821. Solutions
  for NR to support non-terrestrial networks (NTN) (Release 16)}},'' Dec. 2019.

\bibitem{Palacios2022}
J.~Palacios \emph{et~al.}, ``{A Dynamic Codebook Design for Analog Beamforming
  in MIMO LEO Satellite Communications},'' in \emph{Proc. of ICC}, 2022, pp.
  1--5.

\bibitem{PalaciosLEO21}
------, ``{A hybrid beamforming design for massive MIMO LEO satellite
  communications},'' \emph{Frontiers in Space Technologies}, vol.~2, p.~4,
  2021.

\bibitem{Khairallah2021}
N.~Khairallah and Z.~Kassas, ``{Ephemeris Closed-Loop Tracking of LEO
  Satellites with Pseudorange and Doppler Measurements},'' in \emph{Proc. of
  ION GNSS+}, 2021, pp. 2544--2555.

\bibitem{Angeletti2020}
P.~{Angeletti} and R.~{De Gaudenzi}, ``{A Pragmatic Approach to Massive MIMO
  for Broadband Communication Satellites},'' \emph{IEEE Access}, vol.~8, pp.
  132\,212--132\,236, 2020.

\bibitem{You2020}
L.~{You} \emph{et~al.}, ``{Massive MIMO Transmission for LEO Satellite
  Communications},'' \emph{IEEE J. Sel. Areas Commun.}, vol.~38, no.~8, pp.
  1851--1865, 2020.

\bibitem{Wu2005}
W.~Wu, ``{Blockage Mitigation Techniques in Satellite Communications},''
  \emph{IEEE Wireless Commun.}, vol.~12, no.~5, pp. 10--13, 2005.

\bibitem{DiRenzo2020}
D.~Renzo \emph{et~al.}, ``{Smart Radio Environments Empowered by Reconfigurable
  Intelligent Surfaces: How It Works, State of Research, and The Road Ahead},''
  \emph{IEEE J. Sel. Areas Commun.}, vol.~38, no.~11, pp. 2450--2525, 2020.

\bibitem{GEO}
\BIBentryALTinterwordspacing
W.~U. Khan \emph{et~al.}, ``{When RIS Meets GEO Satellite Communications: A New
  Sustainable Optimization Framework in 6G},'' 2022. [Online]. Available:
  \url{https://arxiv.org/abs/2202.00497}
\BIBentrySTDinterwordspacing

\bibitem{Matthiesen21}
B.~Matthiesen \emph{et~al.}, ``{Intelligent Reflecting Surface Operation Under
  Predictable Receiver Mobility: A Continuous Time Propagation Model},''
  \emph{IEEE Wireless Commun. Lett.}, vol.~10, no.~2, pp. 216--220, 2021.

\bibitem{Tang21}
W.~Tang \emph{et~al.}, ``{Wireless Communications With Reconfigurable
  Intelligent Surface: Path Loss Modeling and Experimental Measurement},''
  \emph{IEEE Trans. Wireless Commun.}, vol.~20, no.~1, pp. 421--439, 2021.

\bibitem{Zhang22TWC}
H.~Zhang \emph{et~al.}, ``{Beam Focusing for Near-Field Multi-User MIMO
  Communications},'' \emph{IEEE Trans. Wireless Commun.}, pp. 1--1, 2022.

\end{thebibliography}

\end{document}